\documentclass[byrevtex,showpacs,showkeys,12pt,tightenlines]{revtex4}
\usepackage{amsfonts}
\usepackage{amsmath}
\usepackage{bm}

\newcommand{\Rset}{\mathbb{R}}
\newcommand{\Cset}{\mathbb{C}}
\newcommand{\half}{{\textstyle\frac{1}{2}}}

\newtheorem{thm}{Theorem}[section]

\newtheorem{lem}[thm]{Lemma}
\newtheorem{prop}[thm]{Proposition}
\newtheorem{defn}[thm]{Definition}

\newenvironment{proof}{
\textit{Proof. }\enspace\ignorespaces}{\hbox{}\nobreak\hfill
    \quad\hbox{QED}\par}

\newenvironment{proofof}[1]{
\textit{#1 }\enspace\ignorespaces}{\hbox{}\nobreak\hfill
    \quad\hbox{QED}\par}

\renewenvironment{pmatrix}{\left( \hskip -\arraycolsep
    \begin{array}{ccccc}}{\end{array}\hskip -\arraycolsep \right)}

\newcommand{\hreals}{\hat{\Rset}}
\newcommand{\hcnums}{\hat{\Cset}}

\newcommand{\supth}{^{\mathrm{th}}}

\newcommand{\bu}{\mathbf{u}}
\newcommand{\bv}{\mathbf{v}}

\newcommand{\bV}{\mathbf{V}}

\newcommand{\hz}{\hat{z}}

\newcommand{\hA}{\hat{A}}

\newcommand{\hI}{\hat{I}}

\newcommand{\hT}{\hat{T}}

\newcommand{\brT}{\breve{T}}

\newcommand{\brC}{\breve{C}}

\newcommand{\rA}{\mathrm{A}}
\newcommand{\rB}{\mathrm{B}}
\newcommand{\rC}{\mathrm{C}}
\newcommand{\rD}{\mathrm{D}}

\newcommand{\sA}{{\scriptscriptstyle \rA}}
\newcommand{\sB}{{\scriptscriptstyle \rB}}
\newcommand{\sC}{{\scriptscriptstyle \rC}}
\newcommand{\sD}{{\scriptscriptstyle \rD}}

\newcommand{\du}[2]{_{#1}^{\;#2}}
\newcommand{\ud}[2]{^{#1}{}_{#2}}

\newcommand{\spCK}{\mathbb{CK}}
\newcommand{\spPK}{\mathbb{PK}}
\newcommand{\spCPK}{\mathbb{CPK}}

\newcommand{\spRPK}{\Rset\mathbb{PK}}
\newcommand{\spS}{\mathbb{S}}

\newcommand{\lp}{\left(}
\newcommand{\rp}{\right)}

\newcommand{\tD}{\mathbf{D}}

\newcommand{\tT}{\mathbf{T}}

\newcommand{\pv}{\tilde{v}}
\newcommand{\pT}{\tilde{T}}
\newcommand{\pC}{\tilde{C}}
\newcommand{\pK}{\tilde{K}}
\newcommand{\pR}{\tilde{R}}
\newcommand{\pPhi}{\phi}

\newcommand{\vu}{\mathbf{u}}
\newcommand{\vv}{\mathbf{v}}
\newcommand{\vk}{\bm{\mathit{k}}}
\newcommand{\vell}{{\bm{\ell}}}
\newcommand{\vm}{{\bm{\mathit{m}}}}
\newcommand{\vn}{{\bm{\mathit{n}}}}

\newcommand{\tell}{\tilde{\ell}}

\newcommand{\hvell}{\hat{\vell}}
\newcommand{\hvm}{\hat{\vm}}
\newcommand{\hvn}{\hat{\vn}}

\newcommand{\brell}{\breve{\ell}}
\newcommand{\brn}{\breve{n}}
\newcommand{\brm}{\breve{m}}

\newcommand{\brvell}{\breve{\vell}}
\newcommand{\brvm}{\breve{\vm}}
\newcommand{\brvn}{\breve{\vn}}

\newcommand{\tvell}{\tilde{\vell}}
\newcommand{\tvm}{\tilde{\vm}}

\newcommand{\bwt}{\mathrm{b}}
\newcommand{\bo}{\bwt}
\newcommand{\maxbo}{\bwt_{\max}}

\newcommand{\AS}{\mathcal{A}}

\newcommand{\cW}{\mathcal{W}}

\newcommand{\fvec}[4]{{#1}_{\{a}{#2}_b{#3}_c{#4}_{d\}}}

\newcommand{\bemph}[1]{{\bf #1}}

\newcommand{\lmlm}[2]{\fvec{\ell}{m^{#1}}{\,\ell}{m^{#2}}}

\newcommand{\absbody}{ We develop a dimension-independent theory of
  alignment in Lorentz\-ian geometry, and apply it to the tensor
  classification problem for the Weyl and Ricci tensors.  First,
  we show that the alignment condition is equivalent to the
  PND equation.  In 4D, this recovers the usual Petrov
  types. For higher dimensions, we prove that, in general,
  a Weyl tensor does not possess aligned directions. We then go on to
  describe a number of additional algebraic types for the various
  alignment configurations.  For the case of second-order symmetric
  (Ricci) tensors, we perform the classification by considering the
  geometric properties of the corresponding alignment variety.}
\begin{document}
\title{Alignment and
  algebraically special tensors in Lorentzian geometry.}  
\author{R. Milson}
\email{milson@mathstat.dal.ca}
\author{A. Coley}
\email{aac@mathstat.dal.ca}
\affiliation{%
  Department of Mathematics and Statistics\\
  Dalhousie University\\
  Halifax, Nova Scotia, B3H 3J5 \\
  Canada}
\author{V. Pravda}
\email{pravda@math.cas.cz}
\author{A. Pravdov\'a}
\email{pravdova@math.cas.cz}
\affiliation{%
 Mathematical Institute\\
  Academy of Sciences\\
  \v Zitn\' a 25, 115 67 \\
  Prague 1, Czech Republic
}

\keywords{Alignment, Tensor classification}
\pacs{02.10.Rn ,  04.20.Jb , 04.50.+h}   

\begin{abstract} \absbody \end{abstract}
\maketitle
\section{Introduction}
\label{sect:intro}
In this paper we study the notion of alignment in Lorentz\-ian geometry,
and apply our results to the problem of tensor classification.  Our
ideas are inspired by and generalize covariant classification methods
that are utilized in general relativity.  In four dimensions, tensor
classification is important for physical applications and, in
particular, for the study of exact solutions of the Einstein equations
\cite{kramer}.  Beyond the classical theory, there has been great
interest in higher dimensional Lorentz manifolds as models for
generalized field theories that incorporate gravity \cite{wesson}.
The problem of tensor classification in higher dimensions
\cite{cmpppz,desmet} is therefore of interest.

The most important algebraic classification results in
four-dimensional (4D) general relativity are the classification of the
energy-momentum tensor according to Segre type, and the classification
of the Weyl tensor according to Petrov type.  Our motivation is a
general scheme applicable to the classification of arbitrary tensor
types in arbitrarily high dimensions.  Thus, after developing a
general theory of alignment, we apply our results to the
classification problem of higher-dimensional Weyl tensors.  We also
use alignment to re-derive and extend classification results for
second-order symmetric tensors.

Classification of algebraic tensor types in Lorentzian geometry has a
purely mathematical importance.  The indefinite signature makes
Lorentzian geometry profoundly different from Riemannian geometry.
For example, a second-order symmetric tensor need not be
diagonalizable when the signature is indefinite.  Another striking
difference was recently brought to light in a classification of
Lorentzian 4D manifolds characterized by the property that all scalar
curvature invariants vanish (VSI spacetimes)\cite{cppm}.  In
Riemannian geometry, such a manifold must necessarily be flat, but
this is not so in Lorentzian geometry, where there is a wealth of
non-flat examples.  Thus, understanding of algebraic tensor types in
Lorentzian geometry is needed for the continued investigation of
Lorentzian geometric phenomena, without analogues in Riemannian
geometry.

In practice, a complete tensor classification is possible only for
simple symmetry types and for small dimensions $n$.  However, partial
classification into broader categories is still desirable.  One wants
a classification framework that is as general as possible but, at the
same time, sufficiently detailed so as to reproduce prior
classifications of specific symmetry types and dimensions.

Penrose and Rindler\cite[Ch. 8]{penrind} take a step in this
direction by describing a framework for the classification of
maximally symmetric tensors and spinors in four dimensions.  Their
classification works by associating an algebraic curve on
2-dimensional extended space (the 2-sphere) to a given symmetric
spinor, and then utilizing covariant, algebro-geometric properties of
the curve, such as factor multiplicities and singular points, to
classify the spinor.

The point of departure for the present article is the observation that
the Penrose-Rindler approach can be reinterpreted and generalized in
terms of frame fixing.  The components of a Lor\-entz\-ian tensor can
be naturally ordered according to boost weight.  In essence, one
counts the $\vn$ of the NP tetrad with weight $1$, the $\vell$ with
weight $-1$, and the space-like components with weight $0$.  We will
call a null direction $\vell$ aligned with the tensor, if the
components with the largest weight vanish along that direction. The
Penrose-Rindler curve is just the locus of aligned null directions.
Singular points of the curve correspond to null directions of higher
alignment, ones for which multiple leading orders vanish.

Thus, the key concept in our theory is the notion of an aligned null
direction and of alignment order.  We therefore begin by showing that
the set of null directions aligned with a fixed $n$-dimensional tensor
is a variety (we call it the top alignment variety), meaning that it
can be described by a certain set of polynomial equations in $n-2$
variables.  Covariance now becomes a key issue, because these
alignment equations are given with respect to a particular null-frame.
A change of null-frame transforms the equations in a covariant
fashion.  The necessary mathematical framework needed to describe such
covariant compatibility is a scheme, a very general notion from
algebraic geometry that we adapt to the present circumstances.

The null directions of higher alignment order are subvarieties of the
top alignment variety.  These higher-order directions have an important
geometric characterization related to singularities.  Indeed, we prove
that for irreducible representations of the Lorentz group, the
equations for higher order alignment are equivalent to the equations
for the singular points of the top alignment variety.  Thus, for
tensors, such as trace-free symmetric $R_{ab}$, bivectors, and Weyl
tensors --- these all belong to irreducible representations --- the
algebraically special tensors can be characterized as the instances
whose alignment variety is singular.

Having dispensed with the necessary mathematical preliminaries, we
describe a covariant classification methodology based on alignment
order. In essence, we fix a null-frame $\vell, \vn, \vm_i$ so that the
order of alignment of $\vell$ and $\vn$ along the given tensor is as
large as possible.  We then define the alignment type of a tensor to
be the alignment order of such $\vell$ and $\vn$.  

We illustrate the alignment classification with examples of vectors,
four-dimen\-sional bivectors, second-order symmetric tensors, and
Weyl-like, valence four tensors.  We show how to re-derive the known,
four-dimensional classification results for these tensors --- the
Segre and Petrov types.  We also consider the classification of these
tensor types in higher-dimensions.

For symmetric $R_{ab}$, the higher dimensional classification is a
straight-forward generalization of the situation in four dimensions.
The situation for higher-dimensional Weyl tensors is more complicated.
We prove that the well-known homogeneous PND
equations\cite{penrose60,kramer}, generalized to arbitrary dimensions,
are equivalent to the equations of alignment.  However, we prove that
for $n>4$, these equations do not, generically, possess a solution.
Thus, in contrast to 4D where an aligned $\vell$ and an aligned $\vn$
can always be found, in higher dimensions it is necessary to introduce
new Weyl types to account for the possibility that there does not
exist an aligned $\vell, \vn$.  The end result of this analysis is a
coarse classification, in the sense that it is fully equivalent to the
Petrov classification in four dimension, but does not always lead to
canonical forms for higher-dimensional Weyl tensors.

\section{Preliminaries.}
\subsection{Lorentz and M\"obius geometry.}
\label{sect:geom}
Our setting is $n$-dimensional Min\-kow\-ski space. We define this to
be a vector space isomorphic to $\Rset^n$, together with a
Lorentz-signature inner product $g_{ab}$.  We define a \emph{null
  frame} to be a basis $\vell=\vm_{0}$, $\vn=\vm_{1}$,
$\vm_2,\ldots,\vm_{n-1}$, satisfying $\ell^a n_a = 1$, $m_i{}^a
m_j{}_a = \delta_{ij}$, with all other products vanishing.
In accordance with the above signature convention,
a vector $\bu$ will be called
space-like, time-like, or null depending on whether the norm
$u^a u_a$ is respectively, positive, negative, or zero.

Throughout, roman indices $a,b,c, \rA, \rB, \rC$ range from $0$ to
$n-1$.  Lower case indices indicate an arbitrary basis, while the
upper-case ones indicate a null frame.  Space-like indices $i,j,k$
also indicate a null-frame, but vary from $2$ to $n-1$ only.  
We raise and lower the space-like indices using $\delta_{ij}$, so that
$\vm_i = \vm^i$.  The Einstein summation convention is
observed throughout.

We let $\Phi\ud\sA\sB$ denote a null-frame orthogonal matrix,
and characterize a Lorentz transformation  as a
change of null-frame,  $\hvm_\sB = \vm_\sA \Phi\ud\sA\sB$.
The group of orthochronous\cite{carmeli} Lorentz transformations is generated
by null rotations \eqref{eq:nullrot}, boosts \eqref{eq:boost}, and
spins \eqref{eq:spin}, which are transformations of form
\begin{eqnarray}
  \label{eq:nullrot}
  \hvell&=& \vell
    +z^j \vm_j-\half z^j z_j\, \vn,\quad
    \hvn=  \vn,\quad
    \hvm_i=  \vm_i - z_i \vn;\\
  \label{eq:boost}
  \hvell&=&  \lambda\, \vell, \quad 
  \hvn= \lambda^{-1}\vn,\quad
  \hvm_i=  \vm_i,\quad\lambda \neq 0;\\
  \label{eq:spin}
\hvell&=&  \vell,\quad \hvn= \vn, \quad
  \hvm_j=  \vm_i\, X\ud{i}{j},\quad X\ud{i}{j} X\du{k}{j} =
  \delta\ud{i}{k}.
\end{eqnarray}
The following matrix represents a null-rotation about $\vn$:
\begin{equation}
  \label{eq:nrxform}
  \Lambda\ud\sA\sB(z_i)=
  \begin{pmatrix}
    1 & 0 & 0 \\
    -\half z^j z_j & 1 & -z_i \\
    z^j & 0 & \delta\ud{j}{i}
  \end{pmatrix}
\end{equation}

The set of null lines in $n$-dimensional Minkowski space is an
$m$-dimensional variety
\[
\spPK^m=\{[\vk]: k^a k_a=2 k_0 k_1 + k_i k^i=0\}, \quad m=n-2.
\]
Affine coordinates $z_i = k_i/k_1$ are defined for every choice of
null-frame.  Over the real field, we regard $[\vn]$ as a point at
infinity, and identify $\spRPK^m$ with real extended space
$\hreals^m=\Rset^m \cup \{\infty\}$, the one point compactification of
$\Rset^m$ homeomorphic to the sphere $\spS^m$. We will also need to
consider the complexification of extended space
$$\hcnums^m = \spCPK^m=\Cset^m \cup \spCK^{m-1} \cup \hcnums^{m-2},$$
where
the elements of the second term and third terms are points at infinity
having, respectively, the form $[z^i \vm_i+\vn]$, and $[z^i\vm_i]$,
with $z^i z_i=0$. 

A Lorentz transformation $\hvm_\sB = \vm_\sA \Phi\ud\sA\sB$ induces a
birational transformation of extended space, described by
\begin{equation}
  \label{eq:mobxform}
  z^j = \frac{\pPhi^j(\hz_i)}{\pPhi^0(\hz_i)},\quad 
    \pPhi^\sA(\hz_i) = \Phi\ud{\sA}0   +
  \Phi\ud\sA{i}\,\hz^i-\half\Phi\ud\sA1\, \hz^i \hz_i.
\end{equation}
This can be seen by noting that
\begin{equation}
  \label{eq:framerel}
  \hvell - \half \hz^i \hz_i \hvn + \hz^i \hvm_i =
  \pPhi^0(\hz_i) \vell   + \pPhi^1(\hz_i)\vn  + \pPhi^j(\hz_i) \vm_j\, .
\end{equation}

A real transformation of form \eqref{eq:mobxform} is a conformal
transformation of $\spS^m$, and is known as a M\"obius transformation
\cite{beardon,jeromin}. In terms of the affine coordinates, null
rotations about $\vn$ correspond to translations; null rotations about
$\vell$ to inversions; boosts correspond to dilations; spins
correspond to rotations.  The infinitesimal generators for these
transformations are listed in Table \ref{tab:infgen}.
\begin{table}[h]
  \small
    \caption{infinitesimal generators}
    \label{tab:infgen}
    \begin{tabular}{ccc}
      Generator & Lorentz transformation & M\"obius transformation \\
      \hline 
      \vbox to 12pt{} $\partial_i$ & null rotation about
      $\vn$  & 
      translation \\[3pt]
      $z^j \partial_j $ & boost & dilation \\[3pt]
      $z^i \partial_j- z^j \partial_i$ & spin & rotation\\[3pt]
      $z^i z^j \partial_j - \half z^jz_j \partial_i$ & null
      rotation about $\vell$ & inversion \\[3pt]
      \hline \\
    \end{tabular}
\end{table}

\subsection{M\"obius schemes and subvarieties.}
Let $\Cset[z_i]$ denote the ring of polynomials in the $m$
indeterminates $z_i$. For an affine ideal, $I\subset\Cset[z_i]$, we
let
\[\bV=\bV(I)=\{ \zeta^i\in\Cset^m : p(\zeta^i)=0,\quad p(z^i)\in
I\}\] denote the corresponding affine variety.  Tensor components are
defined relative to a particular choice of null-frame, and transform
covariantly with respect to \eqref{eq:scalarxform}. In order to treat
varieties embedded in extended space, we need to define a similar
notion of covariance for affine ideals.

For $q(z_i)\in\Cset[z_i]$, the fraction ring $\Cset[z_i,q^{-1}] = \{
p(z_i)/q(z_i)^n : p\in\Cset[z_i]\}$ is called the \emph{localization}
by $q$. It represents restriction of a polynomial equations to the
domain $q(z_i)\neq 0$.  We let $I_q = \{
p(z_i)/q(z_i)^d : p\in I\}$ denote the corresponding localization of
an affine ideal $I$.

Let $\hvm_\sB = \vm_\sA \Phi\ud\sA\sB$ be null-frames related by a
Lorentz transformation. Let $\Psi\ud\sA\sB$ be the inverse
transformation,  and let
\begin{equation}
  \label{eq:acchange}
  z^j = \frac{\phi^j(\hz_i)}{\phi^0(\hz_i)},\quad
  \hz^j = \frac{\psi^j(z_i)}{\psi^0(z_i)},
\end{equation}
be the corresponding inverse M\"obius transformations.  
\begin{defn}
\label{defn:idealcov}
We say that affine ideals
$I\subset\Cset[z_i],\;\hI\subset\Cset[\hz_i]$ are \emph{covariantly
  compatible} if the localized ideals $I_{\phi_0}$, $\hI_{\psi^0}$ are
related by the substitutions \eqref{eq:acchange}.  We define a
\emph{M\"obius scheme} $\AS$ to be a covariantly compatible assignment
of an affine ideal $A\subset\Cset[z_i]$ to every null frame $\vm_\sA$.
\end{defn}
We should note that our notion of a scheme is a simplified adaptation
of the usual definition of such objects from algebraic
geometry\cite{eishar}.  

\begin{defn}
We define $\bV(\AS)\subset \hcnums^m$, the
\emph{M\"obius subvariety} corresponding to $\AS$, to be the union of
affine varieties
\[ \bV(\AS)=
\bigcup \{ [\vell - \half \zeta^i \zeta_i \vn + \zeta^i
\vm_i]\in\hcnums^m: p(\zeta_i)=0,\mbox{ for all } p(z_i)\in A\},
\]
where the union is taken over all possible null frames $\vm_\sA$.
\end{defn}

\section{Alignment}
\label{sect:alignment}
\subsection{Boost order.}
Let $\tT=T_{a_1\ldots a_p}$ be a rank $p$ tensor.  For a given list of
frame indices $\rA_1,\ldots,\rA_p$, we call the corresponding
$T_{\sA_1\ldots \sA_p}$ a \emph{null-frame scalar}.  
A Lorentz transformation $\Phi\ud\sA\sB$ transforms the scalars
according to
\begin{equation}
  \label{eq:scalarxform}
   \hat{T}_{\sB_1\ldots \sB_p}=
   T_{\sA_1\ldots \sA_p} \,
   \Phi\ud{\sA_1}{\!\sB_1} \cdots \Phi\ud{\sA_p}{\!\sB_p}.
\end{equation}
In particular, a boost \eqref{eq:boost}  transforms the scalars according to:
\begin{equation}
  \label{eq:boostxform}
  \hat{T}_{\sA_1\ldots \sA_p}= \lambda^{\bwt_{\sA_1\ldots\sA_p}}\,
  T_{\sA_1\ldots \sA_p},\quad  \bwt_{\sA_1\ldots\sA_p}=\bwt_{\sA_1}+\ldots+\bwt_{\sA_p},  
\end{equation}
where   $\bwt_0=1$, $\bwt_i=0$, $\bwt_1=-1$.
\begin{defn}
  We will call $\bwt_{\sA_1\ldots\sA_p}$ \emph{the boost weight} of
  the scalar $T_{\sA_1\ldots\sA_p}$.  Equivalently, the boost weight
  of $T_{\sA_1\ldots \sA_p}$ is the difference between the number of
  subscripts equal to $0$ and the number of subscripts equal to $1$.
  We define the \emph{boost order} of the tensor $\tT$, as a whole, to
  be the boost weight of its leading term, or to put it another way,
  the maximum $\bwt_{\sA_1\ldots\sA_p}$ for all non-vanishing
  $T_{\sA_1\ldots\sA_p}$.
\end{defn}
  
\begin{prop}
Let $\vell,\vn,\vm_i$ and $\hat{\vell},\hat{\vn},\hat{\vm}_i$ be two
null-frames with $\vell$ and $\hat{\vell}$ scalar multiples of each
other.  Then, the boost order of a given tensor is the same relative
to both frames.
\end{prop}
\begin{proof}
  A null rotation about $\vell$ fixes the leading terms of a tensor,
  while boosts and spins \eqref{eq:boostxform} \eqref{eq:spin} subject
  the leading terms to an invertible transformation.  It follows that
  the boost order of a tensor, as a whole, does not depend on a choice
  of a particular null-frame, but rather on the choice of $\vell$. 
\end{proof}

\begin{defn}
  Let $\tT$ be a tensor, and $[\vk],\; k^a k_a=0$ a null direction.
  We choose a null frame $\vell, \vn,\vm_i$, such that $\vell$ is a
  scalar multiple of $\vk$, and define $\bo_\tT(\vk)$, the \emph{boost order
  along $\vk$}, to be the boost order of $\tT$ relative to this frame.
\end{defn}

This definition is sound, because, by the preceding Proposition, the
boost order is the same for all such frames. Usually, the choice of
tensor $\tT$ is clear from the context, and so we suppress the subscript
and simply write $\bo(\vk)$.

\begin{defn}
\label{defn:aligned}
We let $\maxbo$ denote the maximum value of $\bo(\vk)$ taken over all
null vectors $\vk$, and say that a null vector $\vk$ is \emph{aligned}
with the tensor $\tT$ whenever $\bo(\vk)< \maxbo$.  We will call the
integer $\maxbo-\bo(\vk)-1\geq 0$
the order of the alignment.  
\end{defn}

The value of $\maxbo$ depends on the rank and on the symmetry
properties of the tensor $\tT$.  Generically, for a rank $p$ tensor we
have $\maxbo=p$.  However, if the tensor has some skew-symmetry, then
$\maxbo$ will be smaller than $p$.  For example, the boost weights for
a second-order, symmetric tensor $R_{ab}=R_{ba}$ are shown below.
Thus, $\bo(\vell)=1$ if $R_{00}=0$ but some $R_{0i}\neq 0$; while
$\bo(\vell)=0$ if $R_{00}=R_{0i}=0$, and one of $R_{10}, R_{ij}\neq
0$; etc.
\[
  \hskip -2em R_{ab}=\overbrace{R_{00}\, n_a n_b}^2 + 
  \overbrace{2R_{0i} \,n_{(a} m^i{}_{b)}}^1+
      \overbrace{2R_{01}\,\ell_{(a} n_{b)} +
    R_{ij} \,m^i{}_a m^j{}_b}^0 +
  \overbrace{2R_{1i} \,\ell_{(a} m^i{}_{b)}}^{-1} + \overbrace{R_{11}\, \ell_a \ell_b}^{-2}.
\]

For a bivector $K_{ab} = -K_{ba},$ we have $\maxbo=1$; the
corresponding boost weights are shown below. If $K_{0i}=0$, then
$\vell$ is aligned.  If in addition, $K_{01} = K_{ij}=0$, then the
order of alignment is 1.
\begin{equation}
  \label{eq:bvbw}
  K_{ab} = \overbrace{2K_{0i}\, n_{[a} m^i{}_{b]}}^1 +
  \overbrace{2K_{01}\, n_{[a}\ell_{b]}+ K_{ij}\, m^i{}_{[a}
    m^j{}_{b]}}^0+
  \overbrace{2K_{1i}\, \ell_{[a} m^i{}_{b]}}^{-1}.  
\end{equation}

\subsection{The alignment variety}
We now show that the set of aligned directions is a M\"obius variety
by exhibiting a set of compatible equations for this set.  Let $\tT$
be a rank $p$ tensor and $\vm_\sA$ a null-frame.  For every choice of
indices $\rA_1,\ldots,\rA_p$ we define the polynomial
\begin{equation}
  \label{eq:nrpoly}
  \pT_{\sA_1\ldots\sA_p}(z_i) = T_{\sB_1\ldots \sB_p}\,
  \Lambda\ud{\sB_1}{\!\!\sA_1}(z_i)\cdots \Lambda\ud{\sB_p}{\!\!\sA_p}(z_i).  
\end{equation}
where $\Lambda\ud\sB\sA(z_i)$ is the null rotation matrix, defined in
\eqref{eq:nrxform},  parameterized by complex indeterminates $z_i$.

By Definition \ref{defn:aligned}, a null vector $\vell - \half \zeta^i
\zeta_i\vn +\zeta^i \vm_i$, is aligned with $\tT$, with alignment
order $q$, if and only if $z_i=\zeta_i$ satisfies the corresponding
$q\supth$ order \emph{alignment equations}
\begin{equation}
  \label{eq:aligneqs}
  \pT_{\sA_1\ldots\sA_p}(z_i)=0,\quad
  \bwt_{\sA_1\ldots\sA_p}= \maxbo-r,\quad r=0,1,\ldots,q.
\end{equation}  
We call the ideal generated by the above polynomials,
\[A^q=A^q(T)=\langle \pT_{\sA_1\ldots\sA_p}(z_i) \rangle,\quad 
  \bwt_{\sA_1\ldots\sA_p}\geq \maxbo-q,\]
the \emph{alignment ideal} of order $q$.
\begin{thm}
  \label{thm:align}
  Let $\hvm_\sB=\vm_\sA\Phi\ud\sA\sB$ be two complex null frames related by
  Lorentz transformation. Then, the corresponding alignment ideals
  $A^q, \hA^q$ are covariantly compatible in the sense of Definition
  \ref{defn:idealcov}.
\end{thm}
\begin{proof}
  Let $\hT_{\sB_1\ldots\sB_p}$ be the transformed scalars, as per
  \eqref{eq:scalarxform},  and let
  \begin{equation}
    \label{eq:tnrpoly}
    \brT_{\sA_1\ldots\sA_p}(\hz_i) =
    \hT_{\sB_1\ldots \sB_p}\,
    \Lambda\ud{\sB_1}{\!\!\sA_1}(\hz_i)\cdots
    \Lambda\ud{\sB_p}{\!\!\sA_p}(\hz_i) 
  \end{equation}
  be the generators of $\hA^q$.  Effecting substitutions
  \eqref{eq:acchange} as necessary, we have 
  \begin{equation}
    \label{eq:upsrel}
    \Phi\ud\sC\sB \Lambda\ud\sB\sA(\hz_i) = \Lambda\ud\sC\sB(z_i)
    \Upsilon\ud\sB\sA(\hz_i),
  \end{equation}
  where
  \begin{equation}
    \label{eq:upsilondef}
    \Upsilon\ud\sB\sA(\hz_i) = 
    \begin{pmatrix}
      \phi^0 & \Phi\ud01  & \phi\ud0{,i} \\
      0 & \displaystyle \frac{1}{\phi^0} & 0 \\
      0 & \quad \displaystyle \Phi\ud{j}{1} - \Phi\ud01
      \frac{\phi^j}{\phi^0}\quad & \displaystyle \phi\ud{j}{,i} -
      \phi\ud{0}{,i} \frac{\phi^j}{\phi^0}
    \end{pmatrix},
  \end{equation}
  and where
  $\phi^\sA = \phi^\sA(\hz_i)$ as per eq. \eqref{eq:mobxform}.

  To prove that $\Upsilon\ud{\sB}{\sA}$ has the form show in
  \eqref{eq:upsilondef}, we set $\brvm_\sA = \hvm_\sB\,
  \Lambda\ud\sB\sA(\hz_i)$, $\tvm_\sA = \vm_\sB\,
  \Lambda\ud\sB\sA(z_i),$ and note that \eqref{eq:upsrel} is equivalent
  to $\brvm_\sA = \tvm_\sB \Upsilon^\sB_\sA.$ By \eqref{eq:framerel},
  we have that $\brvell = \phi^0\, \tvell$, giving us the first column
  of $\Upsilon^\sB_\sA$.  To verify the form of the second column, we
  note that $\brvn=\hvn$ and use the fact that
  \[ \brn^a \tell_a  = \frac{1}{\phi^0}\, \brn^a \brell_a =
  \frac{1}{\phi^0}.\] By \eqref{eq:mobxform} we have that $\brvm_j =
  \vm_\sA \phi^\sA{}_{,j}(\hz_i)$.  Using this, as well as the fact
  that
  \[ \brm_j{}^a \tell_a = \frac{1}{\phi^0}\, \brm_j{}^a \brell_a = 0, \]
  gives the form of the third column.
    
  Returning to the proof of the Proposition, from
  \eqref{eq:scalarxform} \eqref{eq:nrpoly}
  \eqref{eq:tnrpoly} we have
  \begin{equation}
    \label{eq:eqcovariance}
    \brT_{\sB_1\ldots\sB_p} = \pT_{\sA_1\ldots\sA_p}
    \Upsilon\ud{\sA_1}{\sB_1}  \cdots
    \Upsilon\ud{\sA_p}{\sB_p} \, .
  \end{equation}
  By inspection of \eqref{eq:upsilondef} we see that a given
  $\brT_{\sB_1\ldots\sB_p}$ is a linear combination of $
  \pT_{\sA_1\ldots\sA_p},\; \bwt_{\sA_1\ldots\sA_p} \geq
  \bwt_{\sB_1\ldots\sB_p}$ with coefficients that are polynomials in
  $\Cset[z^i, (\phi^0)^{-1}]$.  It follows that the localization of
  $A^q$ is equal to the localization of~$\hA^q$.  
\end{proof}
\begin{defn}
  We define $\AS^q=\AS^q(\tT)$, the alignment scheme of order $q$, to
  be the scheme generated by the alignment ideals $A^q(\tT)$.  The top
  alignment scheme $\AS^0$, generated by equations
  $\pT_{\sA_1\ldots\sA_p}(z_i)=0,\; \bwt_{\sA_1\ldots\sA_p}= \maxbo$
  has a distinguished role, and will be denoted simply as $\AS=\AS(\tT)$.
\end{defn}

The corresponding varieties $\bV(\AS^q)$ consist of aligned null
directions of alignment order $q$ or more.  It may well happen that
the alignment equations are over-determined and do not admit a
solution, in which case the variety is the empty set. 
\begin{defn}
In those cases
where $\bV(\AS^q)$ consists of only a finite number of null
directions, we will call these directions \emph{principal} and speak
of the \emph{principal null directions} (PNDs) of the tensor.
\end{defn}
\subsection{Singular points and higher alignment.}

\begin{defn}
  Let $I$ be an affine ideal. A point of the variety
  $\zeta_i\in\bV(I)$ will be called \emph{singular} if the first-order
  partial derivatives of the polynomials in $I$ vanish at $\zeta_i$
  also.  It will be called singular of order $q$, if all partials of
  order $q$ and lower, vanish at that point, i.e., $ p_{,j_1\ldots
    j_r}(\zeta_i)=0,\, p(z_i)\in I,\, r\leq q$.
\end{defn}
Geometrically, a singular point represents a self-intersection such as
a a node or a cusp, or a point of higher multiplicity.

\begin{prop}
  Let $\tT$ be a rank $p$ tensor, and $\AS$ the corresponding
  alignment scheme.  Suppose that $\vk = \vell -\half \zeta^i\zeta_i
  \vn + \zeta^i \vm_i$ spans an aligned direction of order $q$, i.e.,
  $z_i=\zeta_i$ satisfies \eqref{eq:aligneqs}. Then, $[\vk]$ is a
  $q\supth$ order singular point of the top variety $\bV(\AS)$.  In
  other words, $z_i=\zeta_i$ also satisfies all equations of the form
  \begin{equation}
    \label{eq:pdaligneqs}
    \pT_{\sA_1 \ldots \sA_p,i_1\ldots i_r}(z_i)=0,\quad 
    \bwt_{\sA_1\ldots\sA_p} = \maxbo,\quad r=0,1,\ldots,q.
  \end{equation}
\end{prop}
\begin{proof}
  From \eqref{eq:nrxform} we have 
  \begin{equation}
    \label{eq:phipartials}
    \Lambda\ud{\sA}{0,i} = \Lambda\ud{\sA}i\quad
    \ \Lambda\ud{\sA}{j,i} = -\delta_{ij}
    \Lambda\ud{\sA}{1},\quad
    \Lambda\ud{\sA}{1,i} = 0.
  \end{equation}
  Hence, by \eqref{eq:nrpoly}
  \begin{equation}
    \label{eq:htpartial}
    \pT_{\sA_1\ldots \sA_p,i} = 
    \pT_{\sB \sA_2 \ldots \sA_p}\lambda\ud{\sB}{\sA_1i} +
    \pT_{\sA_1 \sB 
      \ldots \sA_p}\lambda\ud{\sB}{\sA_2i} + \ldots   +\pT_{\sA_1
    \sA_2 \ldots \sB}\lambda\ud{\sB}{\sA_pi}, 
  \end{equation}  
  where $ \lambda\ud{j}{0i} = \delta\ud{j}{i}$, $\lambda\ud{1}{ji} =
  -\delta_{ij},$ with all other entries zero.  Hence, by
  \eqref{eq:boostxform} the boost weight of the terms in the right hand
  side of \eqref{eq:htpartial} is exactly one smaller than
  $\bwt_{\sA_1\ldots \sA_p}$.  Similarly, every $r\supth$ order
  partial derivative of $\pT_{\sA_1\ldots \sA_p}(z_i)$ is a linear
  combination of polynomials $\pT_{\sB_1\ldots\sB_p}(z_i)$ for which
  $\bwt_{\sB_1\ldots\sB_p}= \bwt_{\sA_1\ldots\sA_p} -r$.  If $[\vk]$
  is aligned, with alignment order $q$, then $z_i=\zeta_i$ satisfies
  \eqref{eq:aligneqs}. Hence, by the preceding remark, $z_i=\zeta_i$
  satisfies \eqref{eq:pdaligneqs} also.  \end{proof} A partial
converse of the preceding is the following.
\begin{prop}
  \label{prop:irrep_sing}
  Suppose that $\tT$ belongs to an irreducible representation of the
  Lorentz group.  Let $\vk = \vell -\half \zeta^i\zeta_i \vn + \zeta^i
  \vm_i$ be a $q\supth$-order singular element of the top alignment
  variety, i.e., $z_i=\zeta_i$ satisfies \eqref{eq:pdaligneqs}. Then,
  $\vk$ spans a $q\supth$ order aligned null direction, i.e.
  $z_i=\zeta_i$ satisfies \eqref{eq:aligneqs} also.
\end{prop}
\begin{proof}
  The polynomials $\pT_{\sA_1\ldots \sA_p}(z_i)$ span an irreducible
  representation of the M\"obius group.  The group action is shown in
  \eqref{eq:eqcovariance}. The corresponding infinitesimal generators
  are matrix differential operators, $\tD + M\ud{\sB}{\sA}$ where
  $\tD$ is an infinitesimal M\"obius transformation, shown in Table
  \ref{tab:infgen}, and where $M\ud{\sB}{\sA}$ is the infinitesimal
  form of $\Upsilon\ud{\sA}{\sB}$. These operators act by
  \[ 
  \hskip -1em \pT_{\sA_1\ldots \sA_p} \mapsto \tD[\pT_{\sA_1\ldots
    \sA_p}] + \pT_{\sB \sA_2 \ldots \sA_p}\, M\ud{\sB}{\sA_1} +
  \pT_{\sA_1 \sB \ldots \sA_p}\, M\ud{\sB}{\sA_2} + \ldots +
  \pT_{\sA_1 \sA_2 \ldots \sB}\, M\ud{\sB}{\sA_p},\] The operators
  corresponding to, respectively, null rotations about $\vn$, boosts,
  spins, and null rotations about $\vell$, have the form
  {\small
  \[
  \partial_i,\quad z^j \partial_j+\begin{pmatrix} 1 & 0 & 0 \\ 0 & -1
  & 0 \\ 0 &0 &0\end{pmatrix},\quad z^i \partial_j- z^j
  \partial_i,\quad
  z^k z^j \partial_j - \half z^jz_j \partial_k +
  \begin{pmatrix}
    -z_k & 0 & -\delta_{ik} \\
    0 & z_k & 0 \\
    0 & \delta\ud{j}{k} & - \delta\ud{j}{k} z_i
  \end{pmatrix}.
  \]}
  The polynomials of maximal boost weight,
  $\pT_{\sA_1\ldots \sA_p}(z_i)$, $\bwt_{\sA_1\ldots \sA_p}=\maxbo,$
  are annihilated by the raising operators and are preserved by the
  infinitesimal spins and boosts.  The lowering operators correspond
  to partial derivatives with respect to the $z_k$. Hence, the partial
  derivatives
  \[
  \pT_{\sA_1  \ldots \sA_p,i_1\ldots i_r}(z_i),\quad 
  \bwt_{\sA_1\ldots\sA_p} = \maxbo,\quad r=0,1,\ldots,2\maxbo
  \]
  span an invariant subspace of the irreducible representation, and
  therefore span the representation.\end{proof}
The preceding two Propositions combine to give the following.
\begin{thm}
  \label{thm:irreps}
  Suppose that $\tT$ belongs to an irreducible representation of the
  Lorentz group.  Then, the $q\supth$ order alignment scheme $\AS^q$
  describes the $q\supth$-order singular points of the top scheme
  $\AS$.  In other words, every equation of form \eqref{eq:aligneqs} is
  a linear combination of equations of form \eqref{eq:pdaligneqs}, and
  vice versa.
\end{thm}

\section{Classification.}
\subsection{Alignment type.}
Algebraically special tensors can be characterized in terms of the
existence of aligned vectors, with increasing specialization indicated
by a higher order of alignment.  In a nutshell, one tries to normalize
the form of the tensor by choosing $\vell$ and $\vn$ so as to induce
the vanishing of the largest possible number of leading and trailing
null-frame scalars. The tensor can then be categorized by the extent
to which such a normalization is possible.

\begin{defn}
  Let $\tT$ be a rank $p$ tensor, and let $\vell$ be an aligned vector
  whose order of alignment is as large as possible. We define the
  \bemph{primary alignment type} of the tensor to be
  $\maxbo-\bwt(\vell)$. If there are no aligned directions, i.e., the
  alignment equations are over-determined, we will say the alignment
  type is G, {\bf general type}.
\end{defn}

Supposing that an aligned $\vell$ does exist, we let $\vn$ be a
null-vector of maximal alignment, but subject to the constraint
$n^al_a=1$. \
\begin{defn}
  We define the \bemph{secondary alignment type} of the tensor to be
  $\maxbo-\bwt(\vn)$, and define the \bemph{alignment type} of the
  tensor to be the pair consisting of the primary and the secondary
  alignment type.
\end{defn}
If $\vell$ is the unique aligned direction,
i.e.  if no aligned $\vn$ exists, then we define the alignment type to
be the singleton consisting of the  primary alignment type. We will
also speak of complex and real types, according to whether or not we
permit $\vell, \vn$ to be complex.


By way of example, let us apply the alignment formalism to the
description of vectors.    Let $\vv = v_0\, \vn  +v_1\,\vell+v_i\, \vm^i$
be a non-zero, real vector.  By \eqref{eq:nrxform} \eqref{eq:nrpoly}
\eqref{eq:aligneqs}, the zeroth and first order alignment equations
are, respectively,
\begin{eqnarray}
  \label{eq:vector1}
  \pv_0 &=& v_0 + z^i v_i  -
  \half z^i z_i v_1 = 0,\\
  \label{eq:vector2}
  \pv_i &=& v_i - z_i v_1 = 0.
\end{eqnarray}
The covariance of the above equations is easily verified using
\eqref{eq:eqcovariance}.  Also note that, as asserted by Theorem
\ref{thm:irreps}, equations \eqref{eq:vector2} are spanned by the
partial derivatives of equation \eqref{eq:vector1}.

Rewriting \eqref{eq:vector1} as
\[
\delta^{ij} (v_1 z_i - v_i) (v_1 z_j - v_j) = 2 v_0 v_1 + v^i v_i,
\]
we see that there are three possibilities, according to
the sign of the norm $v^a v_a$.  If $\bv$ is space-like, i.e., the
norm is positive, then $\bV(\AS)$ is a hypersphere.  If $\bv$ is
null, then $\bV(\AS)$  consists of a unique PND of alignment order $1$.
If $\bv$ is time-like,
then  $\bV(\AS)$ is a complex hypersphere;  there are no real
aligned directions. 

Thus, space-like vectors have alignment type (1,1).  This means that
every space-like vector can be put into the form $\zeta^i \vm_i$.
Time-like vectors are of real type G, but have complex alignment type
(1,1).  They can also be normalized to the form $\zeta^i \vm_i$, but
the required null frame and coefficients $\zeta_i$ need to be complex.
Finally, a null vector $\vk,\; k^a k_a=0$ has alignment type (2).
This is because a null vector can be normalized to the form $\vell$ by
taking a null frame for which $\vell=\vk$.  There is no secondary
type, because a null vector can only be aligned with a multiple of
itself.

\subsection{Four-dimensional bivectors.}
It will be useful, at this point, to consider the covariant
classification of bivectors in four-dimensional Minkowski space.  The
bivector boost weights are shown in \eqref{eq:bvbw}.  Applying
\eqref{eq:nrxform} \eqref{eq:nrpoly} we obtain the following form
for the alignment equations:
\begin{eqnarray}
  \label{eq:4dbv11}
  \pK_{02} &=& K_{02}- K_{01} z_2 - K_{23} z_3 + \half
  K_{12}(z_2^2 -   z_3^2) +K_{13} z_2 z_3=0\,,\\
  \label{eq:4dbv12}
  \pK_{03} &=& K_{03} +K_{23} z_2 - K_{01} z_3  -\half
  K_{13}(z_2^2 -  z_3^2) +K_{12} z_2 z_3 =0\,.
\end{eqnarray}
According to Theorem \ref{thm:irreps}, the equations for first order
alignment can be obtained by taking partial derivatives of
\eqref{eq:4dbv11}\eqref{eq:4dbv12}:
\begin{eqnarray}
  \label{eq:4dbv21}
  \pK_{01} &=& -\pK_{02,2}=  -\pK_{03,3} 
  = K_{01} - K_{12}
  z_2 - K_{13} z_3 =0\,,\\ 
  \label{eq:4dbv22}
  \pK_{23} &=& -\pK_{02,3} =\pK_{03,2}= K_{23} - K_{13} z_2 + K_{12} z_3=0\,.
\end{eqnarray}
Without loss of generality $K_{13}, K_{12}$ are not both zero. By
performing a spin and then a boost we can change to a null-frame where
$K_{13}=0,\quad K_{12}=1$.  The top alignment equations
\eqref{eq:4dbv11}\eqref{eq:4dbv12} may now be written as
\begin{eqnarray}
  (z_2-K_{01})^2 - (z_3 +K_{23})^2 &=& (K_{01})^2 - (K_{23})^2 - 2 K_{02}\,,\\
  (z_2-K_{01})(z_3 +K_{23}) &=& -K_{01}K_{23} - K_{03}\,.
\end{eqnarray}
Generically, these equations have 4 discrete solutions: two real and
two complex.  These correspond to 2 real and 2 complex principal null
directions.  Aligning $\vell$ and $\vn$ gives the canonical form:
$K_{ab} = \lambda\, m^2{}_{[a} m^3{}_{b]} + \mu\, \ell{}_{[a}
n{}_{b]}.  $ Thus, generically, the alignment type is (1,1).  The
exceptional case corresponds to
\[  (K_{01})^2 - (K_{23})^2 - 2 K_{02}=0,\quad K_{01}K_{23}+K_{03}=0\,.\]
Here, the complex PNDs merge with the real ones to create a singular,
real PND.  In this case, the canonical form is
$K_{ab} = l{}_{[a} m^2{}_{b]}$.
Such a bivector has alignment type
(2).

\subsection{Second-order symmetric tensors.}
\label{sect:r2sym}
Let $R_{ab}=R_{ba}$, $R\ud{a}{a}=0$, be a traceless, second-order
symmetric tensor.  The maximum boost weight is $2$, with $R_{00}$ the
scalar of maximum boost weight.  Thus, by \eqref{eq:nrxform}
\eqref{eq:nrpoly} the unique top alignment equation has the form
\begin{equation}
  \label{eq:qf0align} 
  \pR_{00} =
  R_{00} + 2 R_{0i}z^i  + R_{ij}z^i z^j  -  2R_{01}r^2
  - 2R_{1j}  z^j  r^2+  R_{11} r^4=0,
\end{equation}
where we have defined
$r^2 = \half z^i z_i$.
The  homogeneous form of \eqref{eq:qf0align} is
\begin{equation}
  \label{eq:qfhomalign}
  R_{ab}\, k^a k^b = 0,\quad k^a k_a=0.
\end{equation}

Alignment type, by itself, is insufficient to fully classify symmetric
$R_{ab}$.  To obtain a classification, it is necessary to make use of
the algebro-geometric properties of the alignment variety defined by
equation \eqref{eq:qf0align}. 

This approach is a natural generalization of the Penrose-Rindler
method for covariant classification of four-dimensional, symmetric
tensors and spinors \cite{penrind}.  In four dimensions, equations
\eqref{eq:qf0align} \eqref{eq:qfhomalign} describe a certain class of
algebraic curves on the 2-sphere.  A comprehensive classification of
$R_{ab}$ based on the classification of such curves is due to Penrose
\cite{penrose72}.  See \cite[Ch. 5]{kramer} for other classification
approaches and additional references.

For our part, we will describe a generalization of the Penrose
classification to arbitrary dimensions.  The key technique here is the
expression of the alignment equation \eqref{eq:qfhomalign} using the
affine form \eqref{eq:qf0align}.  Generically, \eqref{eq:qf0align}
describes a fourth degree polynomial.  However, as soon as one
considers singular cases, one can perform a M\"obius transformation
and reduce \eqref{eq:qf0align} to a quadratic polynomial, greatly
facilitating the classification.  Indeed, the bulk of the
classification --- the singular cases --- is reduced to the problem of
classifying quadratic equations in $m$ variables up to Euclidean
transformations.

We do this in detail for the 4D case. The higher dimensional
classification, while admitting more particular cases, does not
materially differ from the four-dimensional case.  The canonical forms
for the four-dimensional alignment equations are shown in Table
\ref{tab:4dqform}.  The multiplicities in the Segre type (third
column) are listed in order from lowest to highest eigenvalue.  A
comma is used to separate an eigenvalue with a time-like eigenvector.
The last column shows the alignment type.  Where the complex alignment
type differs from the real one, the complex type is given first.

\begin{table}[h]
  \begin{center}
    \begin{tabular}{rclccc}
      &$\pR_{00}(x,y)$ & & Segre type & Align. type\\ \hline
      \vbox to 12pt{}
 1. &     $r^4 + a x^2 + by^2 +1$ &$,\;1<a<b$ & $111,1$ & (1,1)/G\\
 2. &     $r^4 + a x^2 + by^2 -1$ & $,\;a\neq b$ &$11,Z\overline{Z}$ & (1,1)\\  
 3. &     $r^4 + a x^2 + by^2 +1$ & $,\;a<b<-1$& $11,1,1$ & (1,1)\\
 4. &     $r^4 + 2ar^2 + 1 = 0$ & $,\; a>1 $ & $1,1(11)$ & (2,2)/G\\
 5. &     $r^4 + 2ar^2 + 1 = 0$ & $,\; |a|<1 $ & $1,(11)1$ & (2,2)/G\\
 6. &     $r^4 + 2ar^2 + 1 = 0$ & $,\; a<-1$ & $(11),1,1$
      & (2,2)/(1,1)\\
 7. &     $r^4 + 2ar^2 -1 = 0$ & & $(11),Z\overline{Z}$ & (2,2)/(1,1)\\
 8. &     $(r^2+1)^2$ & & $1,(111)$ & (2,2)/(2)\\
 9. &     $-x^2+ay^2+1$ & $,\;a>0$ & $121$ & (2,1)\\
 10. &     $-x^2+ay^2+1$ & $,\;a<0$ & $112$ & (2,1)\\
 11. &     $x^2+ay^2+1$ & $,\;a>0$ & $211$ & (2,1)/(2)\\
 12. &     $x^2+ay^2$ & $,\;a<0$ & $1(1,1)1$ & (2,2)\\
 13. &     $x^2+ay^2$ & $,\;a>0$ & $(1,1)11$ & (2,2)\\
 14. &     $x^2-y$ &  & $13$ & (2,1)\\
 15. &     $-x^2-y^2+1$ & & $(11)2$ & (2,1) \\
 16. &     $x^2+y^2+1$ & & $2(11)$  & (2,1)/(2)\\
 17. &     $-x^2 +1$ && $1(12)$ & (2,1) \\
 18. &     $x^2 +1$ && $(12)1$ & (2,1)/(2)\\
 19. &     $x^2+y^2$ && $(11)(11)$ & (2,2)\\
 20. &     $x^2$ & & $(11,1)1$ & (2,2)\\
 21. &     $x$ && $(13)$ &(3,1)\\
 22. &     $1$ && $(112)$ & (4)\\
 23. &     $0$ 
    \end{tabular}\smallskip
    \caption{Canonical alignment equations for  $R_{ab}$ in 4D}
    \label{tab:4dqform}
  \end{center}
\end{table}

The first step in the classification, is to separate the singular and
the non-singular cases.
\begin{prop}
  \label{prop:singaqf}
  The alignment variety \eqref{eq:qf0align} is singular if and
  only if $R\ud{a}{b}$ possesses a multiple eigenvalue.
\end{prop}
\begin{lem}
  \label{lem:nullvec}
  The linear transformation $R\ud{a}{b}$ has a multiple eigenvalue if
  and only if it possesses a real or a complex null eigenvector.
\end{lem}
\begin{proof}
  Suppose that $\vn$ is a null eigenvector with eigenvalue $\lambda$.
  We complete to a null frame $\vell, \vn, \vm_i$, and note that
  $R_{11} = R_{1i} =0$, and that $R_{01} = \lambda$.  Let us consider
  two cases, according to whether $\lambda$ is an eigenvalue of the
  matrix $R\ud{i}{j}$.  If yes, let $\bv = v^i \vm_i$ be an
  eigenvector.  Hence,
  \[  R\ud{a}{b}\, v^b = \lambda v^a + R_{0i} v^i n^a,  \]
  which proves
  that $\lambda$ is a multiple eigenvalue.  In the opposite case, the
  matrix $R\ud{i}{j}-R_{01} \delta\ud{i}{j}$ is non-singular, and
  hence by performing a null-rotation about $\vn$ (equivalently, a
  translation in the $z_i$) we can switch to a null frame where
  $R_{0i}=0$.  Now, we have
  \[ R\ud{a}{b}\, \ell^b = \lambda \ell^a + R_{00} \, n^a, \]
  which again shows that $\lambda$ is a multiple eigenvalue.
  
  Conversely, suppose that $R\ud{a}{b}$ has a multiple eigenvalue
  $\lambda$.  Hence there exist vectors $\vu, \vv$ such that
  \begin{eqnarray*}
    R\ud{a}{b} \, u^b &=&  \lambda u^a,\\
    R\ud{a}{b} \, v^b &=& \lambda v^a + \kappa u^a.
  \end{eqnarray*}
    
  Consider the case $\kappa\neq 0$.  We have
  \[ R_{ab} u^a v^b = \lambda\, u^a v_a = \lambda\, v^a u_a +\kappa \,
  u^a u_a, \] and hence $u^a u_a=0$, as desired.  Finally, suppose
  that $\kappa=0$.  If either $\vu$ or $\vv$ are null eigenvectors, we
  are done.  If neither is a null-vector, we construct a null
  eigenvector (possibly complex) of the form $\vu+\alpha \vv,$ where
  $\alpha$ is a solution of the following quadratic equation:
  \[ u^a u_a + 2 u^a v_a\, \alpha +  v^a v_a\, \alpha^2=0.\]  
\end{proof}

\begin{proofof}{Proof of proposition \ref{prop:singaqf}.}
  Suppose that \eqref{eq:qf0align} is singular.  Choose an $\vell$
  that spans a singular direction, and complete to a null frame
  $\vell, \vn, \vm_i$.  Note that if the singular direction is
  complex, then the null frame will have to be complex also.  By
  Theorem \ref{thm:irreps} we have $ R_{00} = R_{0i} =0,$ and hence $
  R\ud{a}{b}\, \ell^b = R_{01} \ell^a.$ Hence, $\vell$ is a null
  eigenvector and, by the Lemma, $R\ud{a}{b}$ has multiple
  eigenvalues.  Conversely, suppose that $R\ud{a}{b}$ has multiple
  eigenvalues.  By the lemma, we can choose a null eigenvector
  $\vell$.  Completing, to a null frame, we obtain $ R_{00} = R_{0i}
  =0,$ which proves that $\vell$ spans a singular direction.
\end{proofof}

Turning first to the non-singular case, by the above Proposition all
eigenvalues are distinct, and hence we can choose an orthonormal set
of $m=n-2$ space-like eigenvectors $\vm_i$ with eigenvalues
$\lambda_i$.  We complete these to a null-frame, and perform a boost
so that $R_{00} = \pm R_{11}$.  In this way the alignment equation
simplifies to the following canonical form
\begin{equation}
  \label{eq:qfnormform}
  r^4 + \sum_{i} a_i z_i{}^2 \pm 1=0,\qquad
  a_i =\frac{ \lambda_i -  R_{01}}{R_{11}}.
\end{equation}
We now distinguish two sub-cases, depending on the sign in the above
equation.  If the sign is negative, the alignment variety is
homeomorphic to the sphere $\spS^{m-1}$.  One can show that this case
is distinguished by the presence of a pair of conjugate complex
eigenvalues of $R\ud{a}{b}$.

If the sign in \eqref{eq:qfnormform} is positive, the topology of the
real alignment variety is determined by the position of the eigenvalue
corresponding to the time-like eigenvector. Let $p$ be the number of
times when $a_i<-1$.  If $p>0$, then the real part of the alignment
variety is homeomorphic to a product of spheres, $\spS^{p-1} \times
\spS^{m-p}$, where we define the zero-dimensional sphere $\spS^0$ to
be the set consisting of two distinct points.  If $p=0$, the alignment
variety has no real points.  In all these cases, the eigenvalues of
$R\ud{a}{b}$ are real, with exactly one eigenvalue corresponding to a
time-like eigenvector.

In 4D, the non-singular cases are described by lines 1, 2, 3 of Table
\ref{tab:4dqform}.  In case 1, the alignment variety has no real
points.  For cases 2 and 3, the alignment variety is homeomorphic to
$\spS^1$ and $\spS^1\times \spS^1$, respectively.

In the degenerate cases, Lemma \ref{lem:nullvec} shows that there
exists at least one (possibly complex) null eigenvector.  The
degenerate cases are, therefore, divided according to whether the null
eigenvector of $R\ud{a}{b}$ is real or complex.  If the null
eigenvector is complex, then the canonical form of the alignment
equation is \eqref{eq:qfnormform}, with $a_i=a_j$ for some $i\neq j$.
In 4D, the singular cases with a complex eigenvector correspond to
lines 4-8 of Table \ref{tab:4dqform}.

If there exists a real null eigenvector, we use a null frame $\vell,
\vn, \vm_i$ with $\vn$ spanning the singular direction.  In this way
$R_{11} = R_{1i} = 0,$ and \eqref{eq:qf0align} reduces to a quadratic
equation.  The subgroup of M\"obius transformations preserving $[\vn]$
is precisely the group of Euclidean similarity transformations:
rotations, translations, and scalings.  Thus, in the real, singular
case the classification reduces to the familiar problem of classifying
conics up to Euclidean transformations.  In 4D, these cases correspond
to lines 9-23 of Table \ref{tab:4dqform}.


\subsection{Weyl-like tensors.}
\label{sect:weyl}
We define a \emph{Weyl-like tensor} $C_{abcd}$ to be a traceless,
valence 4 tensor with the well-known index symmetries of the Riemann
curvature tensor, i.e.,
\[ C_{abcd} = -C_{bacd} = C_{cdab},\quad
C_{abcd}+C_{acdb}+C_{adbc}=0,\quad C_{abc}{}^b=0.\] We let $\cW_n$
denote the vector space of $n$-dimensional Weyl-like tensors.  It
isn't hard to show that $\cW_n$ has dimension
$\frac{1}{12}(n+2)(n+1)n(n-3)$.



\begin{table}[h]
    \caption{Boost weight of  the Weyl scalars.}
    \label{tab:bw2}
    \begin{tabular}{|c|c|c|c|c|}
      \hline
      $2$ & $1$ & $0$ & $-1$ & $-2$\\
      \hline
      $C_{0i0j}$ &  $C_{010i}, C_{0ijk}$ & $C_{0101}, C_{01ij},
      C_{0i1j}, C_{ijkl}$ & $C_{011i}, C_{1ijk}$ & $C_{1i1j}$\\
      \hline
    \end{tabular}
\end{table}


The maximal boost weight for a Weyl tensor is given by $\maxbo=2$.
There are $\half n(n-3)$ independent scalars of maximal boost weight.
We define the Weyl alignment equations to be the top alignment
equations 
\begin{equation}
  \label{eq:weylalign}
  \pC_{0i0j}=0,
\end{equation}
where the left-hand side is defined in
\eqref{eq:nrpoly}.  This is a system of $\half n(n-3)$, fourth order
equations in $n-2$ variables.

In 4D, the principal null directions of the Weyl-like
tensor are defined in terms of the so-called PND
equation\cite{penrose60,kramer}:
\begin{equation}
  \label{eq:pndeq}
 k^b k_{[e} C_{a]bc[d}  k_{f]}k^c=0,\quad k^a k_a=0.
\end{equation}
Let us show that, in all dimensions, the above system of equations is
equivalent to the Weyl alignment equations \eqref{eq:weylalign}
\begin{prop}
  For every dimension $n$, a null vector $\vk$ satisfies
  \eqref{eq:pndeq} if and only if $\vk$ is aligned with $C_{abcd}$.
\end{prop}
\begin{proof}
  Let a null vector $\vk$ be given.  Completing to a null-frame
  $\vell=\vk, \vn, \vm_i$,   we have
  \begin{eqnarray*}
   && k^b C_{abcd} k^c = -C_{0i0j} m\ud{i}{a} m\ud{j}{d} -2C_{010i}
    \ell_{(a} m\ud{i}{b)}-C_{0101} \ell_a \ell_b,\\
    &&k^b k_{[e} C_{a]bc[d}  k_{f]}k^c  =
    -C_{0i0j}\, \ell_{[e} m\ud{i}{a]} \ell_{[f} m\ud{j}{d]}    
  \end{eqnarray*}
  Hence,  \eqref{eq:pndeq} holds if and only if 
  $C_{0i0j}=0$,
  i.e., all scalars of boost weight 2 vanish.  This is precisely the
  definition of $\vk$ being aligned with $C_{abcd}$.  \end{proof}

Henceforth, let us set $m=n-2$. For $n\geq 4$ we have 
\[ \half (m+2)(m-1) \geq m,\]
with equality if and only if $n=4$.  Thus, a four-dimensional
Weyl-like tensor always possesses at least one aligned direction (see
below).  The situation in higher dimensions is described by the
following:
\begin{thm}
  \label{thm:nopnds}
  If $n\geq 5$, then the subset of $\cW_n$ with complex alignment type
  G is a dense, open subset of $\cW_n$. In other words, the generic
  Weyl-like tensor in higher dimensions does not possess any aligned
  null directions (not even complex ones).
\end{thm}
\begin{proof}
  Let $C_{abcd}$ be a Weyl-like tensor with scalars $C_{\sA\sB\sC\sD}$
  relative to some fixed null frame. Let $C'_{abcd}$ be a Weyl-like tensor
  of boost order $1$ defined by
  \[
  C'_{\sA\sB\sC\sD} = 
  \left\{ \begin{array}{ll}
        C_{\sA\sB\sC\sD} & \mbox{ for } \bwt_{\sA\sB\sC\sD} \leq 1 \\
        0 & \mbox{ otherwise.}
      \end{array} \right.
    \]
  In other words, 
  \[\pC_{0i0j}(z_k)  = C_{0i0j}+ \pC'_{0i0j}(z_k).\]
  Next, let us define the mapping $\gamma: \Cset^m \rightarrow
  \Cset^{\frac12 m(m+1)}$ by
  \[\gamma_{ij}(z_k) = -\pC'_{0i0j}(z_k),\quad i\leq j.\]
  Hence,
  $C_{abcd}$ possesses a non-infinite aligned direction if and only if
  $C_{0i0j}$ is in the image of $\gamma$.  However, generically, the
  image of the mapping $\gamma$ is an affine subvariety of dimension
  $m$, whereas the set of all $C_{0i0j}$ satisfying $C_{0i0}{}^i=0$ is
  a subspace of dimension $\half(m+2)(m-1)$.  Therefore, the set of
  all $C_{abcd}$ with an aligned null direction is an $m$-dimensional
  subvariety of the affine space of all Weyl-like tensors.  The
  desired conclusion follows immediately.  \end{proof}

Let us now re-derive the well-known Petrov classification of
4-dimensional Weyl-like tensors in terms of alignment type.  The
calculations are facilitated by the use of the NP
tetrad\cite{penrind}, and so we introduce the following notation:
\begin{eqnarray*}
  z_{\pm} &=& \frac{1}{\sqrt{2}} ( z_2 \pm i z_3), \qquad
  \partial_{\pm} = \frac{1}{\sqrt{2}}\lp \partial_2 \mp i
  \partial_3\rp,\qquad
  \vm_{\pm}= \frac{1}{\sqrt{2}} ( \vm_2 \mp \vm_3)\\
  \Psi &=& \pC_{0202} + i \pC_{0203} ,\qquad
  \Lambda\ud{\sA}{\pm} = \frac{1}{\sqrt{2}} \lp \Lambda\ud{\sA}{2} \pm i
  \Lambda\ud{\sA}{3} \rp,
\end{eqnarray*}
where, $\pC_{\sA\sB\sC\sD}$ is given by \eqref{eq:nrpoly},
and where, as per \eqref{eq:nrxform}, we have
\[
\Lambda\ud{\sA}{\sB} = 
  \begin{pmatrix}
    1 & 0 & 0 & 0\\
    -z_+ z_- & 1 & z_2 & z_3 \\
    -z_2 & 0 & 1 &0\\
    -z_3 & 0 & 0 & 1
  \end{pmatrix}
\]
Rewriting \eqref{eq:phipartials} we have that
\[
  \Lambda\ud{\sA}{0,\pm} =
  -\Lambda\ud{\sA}{\pm},\quad
  \Lambda\ud{\sA}{\pm,\pm} =
  \Lambda\ud{\sA}{1},\quad
  \Lambda\ud{\sA}{\pm,\mp} = 
  \Lambda\ud{\sA}{1,\pm} = 0.
\]
It follows immediately that $\Psi_{,-} = 0.$ Hence,
\[
  \Psi(z_+,z_-)=\Psi(z_+)= (C_{1212}+ i
  C_{1213})(z_+-\zeta_1)(z_+-\zeta_2)(z_+-\zeta_3)(z_+-\zeta_4)
\]
is a fourth degree polynomial of one
complex variable.
Rewriting the top alignment equations as
$\Psi(z_+)=\overline{\Psi}(z_-)=0$,
we deduce that, generically, there  are 16 solutions:
\[
 z_2 = \frac{1}{\sqrt{2}}(\zeta_M + \bar{\zeta}_N),\quad
z_3 = \frac{-i}{\sqrt{2}}(\zeta_M - \bar{\zeta}_N),\quad
M,N=1,2,3,4 .
\]
The 4 real solutions correspond to the case $M=N$.

By Theorem \ref{thm:irreps} (this can also be verified directly) the
equations for higher order alignment are given by the derivatives 
\[
\Psi^{(r)}(z_+)=0,\quad \overline{\Psi^{(r)}}(z_-)=0,\quad
r=0,\ldots,q,\quad q=1,2,3.
\]
It follows that the equations for
alignment order $q$ have a solution if and only if $\Psi(z_+)$
possesses a root of multiplicity $q+1$ or more.  In this way, we
recover the usual Petrov classification, which counts the root
multiplicities of the polynomial $\Psi(z_+)$ --- see Table
\ref{tab:4dpetrov}.    
\begin{table}[h]
    \caption{The 4D Petrov classification in terms of alignment type.}
    \label{tab:4dpetrov}
    \begin{tabular}{l@{\hskip 1em}cccccc}
      \hline
      Weyl type &   I & II & D & III & N & O\\
      Alignment type  & (1,1) & (2,1) & (2,2)& (3,1) & (4) & (5)\\
      root multiplicities of $\Psi$ & 1,1,1,1 & 2,1,1,1& 2,2 & 3,1& 4\\
      \hline
    \end{tabular}
\end{table}


\begin{table}[h]
  \caption{Higher dimensional Weyl types.}
  \label{tab:weylclass}
    \begin{tabular}{l@{\hskip 1em}cccccccccc}
      \hline
      Weyl type &   G& I & $\mathrm{I}_i$ & II & $\mathrm{II}_i$& D &
      III & $\mathrm{III}_i$ & N & O\\
      Alignment type  & G &(1) & (1,1) & (2) & (2,1) & (2,2)& (3) &
      (3,1) & (4) & (5)\\ 
      \hline
    \end{tabular}
\end{table}


Table \ref{tab:weylclass} lists the possible alignment types of a
higher dimensional Weyl-like tensor, grouped into categories which are
compatible with the usual Petrov classification.  By
Theorem~\ref{thm:nopnds}, in 5 dimensions and higher, the generic
Weyl-like tensor does not have any aligned directions. Thus, unlike
the case in 4D, type I tensors are an algebraically special category.

Another key difference in the higher-dimensional classification is the
significance of the secondary alignment type.  The proof of Theorem
\ref{thm:nopnds} is easily adapted to show that, for types I, II, and
III, even though an aligned $\vell$ exists, there does not,
generically, exist an aligned $\vn$.  This is in contrast to 4
dimensions, where an aligned $\vn$ can always be chosen.  Hence, in
higher dimensions we must distinguish types $\mathrm{I}_i,
\mathrm{II}_i, \mathrm{III}_i$ as the algebraically special subclasses
of types I, II, and III that possess an aligned $\vn$.  Type
$\mathrm{II}_{ii}$ is a further specialization of type
$\mathrm{II}_i$, one where $\vn$ has alignment order 1.  It is denoted
by type D in analogy with the 4-dimensional classification.  As
expected, type O is the maximally degenerate type corresponding to the
vanishing of the Weyl-like tensor.

Also, unlike the 4-dimensional classification, the above categories
are coarse in the sense that they collect together a number of
inequivalent tensor types.  Thus, in the 4D classification each
alignment type admits a canonical form, but this is not always true
for  $n>4$.  Section \ref{sect:disc} contains some additional remarks
on the classification of curvature tensors in higher-dimensional
Lorentz spaces.

\section{Discussion}
\label{sect:disc}
The present paper develops the theory of alignment in
Lorentzian geometry.  We have defined and studied the general notions
of \emph{aligned null direction}, \emph{order of alignment}, and
\emph{alignment type}, and applied these ideas to the problem of
tensor classification.  In particular, we have argued that it is
possible to categorize algebraically special tensors in terms of their
alignment type, with increasing specialization indicated by a higher
order of alignment.

Alignment type suffices for the classification of 4-dimensional
bivectors and Weyl-like tensors, but is not sufficiently refined for
the classification of second-order symmetric tensors and the
classification of tensors in higher dimensions.  Thus, the present
paper should be considered as a necessary first step in the
investigation of covariant tensor properties based on the notion of
alignment.  


Of particular interest are classes of tensors corresponding to
irreducible representations of the Lorentz group, because by Theorem
\ref{thm:irreps} the higher alignment equations describe the
singularities of the top alignment variety.  Thus, for irreducible
representations the classification problem is reduced to the study of
the corresponding moduli space --- the variety of all top alignment
schemes quotiented by the group of M\"obius transformations.  The
algebraically special classes correspond to the various singular
strata of this space.


The representation of Weyl-like tensors is of particular interest,
both from the theoretical standpoint and because of physical
applications.  In higher dimensions, classification needs to go beyond
alignment type, and to consider other covariant properties of the Weyl
aligned null directions (WANDs).  The key concepts and theoretical
issues underlying such a classification are sketched out in
\cite{cmpp}, and are also summarized below.

{\bf 1. Orthogonal reducibility.}
Let $\vm_\sA$ be a null-frame.  Let us fix a $k=2,\ldots,n-1$, and partition
the null frame into a null subframe
$\vell, \vn, \vm_2,\ldots,\vm_{k-1}$
and a positive-definite subframe
$\vm_k,\ldots,\vm_{n-1}$.
Having done this, it will be convenient to
call indices $\rA=0,1,\ldots,k-1$ Lorentzian, and call indices
$\rA=k,\ldots,n-1$ Riemannian.  Let $C_{abcd}$ be a Weyl-like tensor.
Scalars $C_{\sA\sB\sC\sD}$ can now be categorized as being
Lorentzian, Riemannian, or mixed, according to whether the indices are
all of one type or the other, or whether there are some indices of
both types.  We will say that the frame $\vm_{\sA}$ defines an
\emph{orthogonal decomposition} and that  $C_{abcd}$ is
\emph{orthogonally reducible} if for some $k=2,\ldots,n-1$ all mixed
scalars vanish.

Orthogonal reducibility is a covariant criterion, and refines the
alignment type categories.  Indeed, suppose that $C_{abcd}$ is
reducible, and let $\brC_{\sA\sB\sC\sD}$ denote the $k$-dimensional
reduction obtained by restricting to Lorentz indices.  We can then
define the \emph{reduced alignment type} of $C_{abcd}$ to be the
alignment type of the reduced tensor.  Uniqueness of the orthogonal
decomposition now becomes a key theoretical question.  A closely
related question is the extent to which the alignment equations of a
reducible tensor decouple into separate equations for the Lorentzian
and the Riemannian variables.

Let us also note that non-trivial Weyl-like tensors do not exist if
$n<4$.  Thus, reducibility becomes an issue only in dimensions $n\geq
5$, and even then, one of the complementary summands must necessarily
vanish for $n\leq 7$.  True decomposability can only occur for $n\geq
8$.

{\bf 2.  Cardinality of aligned null directions.}  For dimension $n=4$
there are exactly four WANDs, counting multiplicities.  For $n\geq 5$,
the situation is more complicated.  If all WANDs are principal, what
is an upper bound for each dimension $n$?  It is also possible to have
an infinity of WANDs, as in the case of an orthogonally reducible
tensor with vanishing Riemannian components.  When does this type of
degeneracy occur, or conversely, what is required for WANDs to be
principal?

{\bf 3. Canonical forms.}  By definition, classification according to
alignment type corresponds to a normal form where the components of
leading and trailing boost weight vanish.  It would be desirable to
refine the classification to the point of a  true canonical form.
This is possible in the case of type N Weyl-like tensors, which can always be
put into the form $\sum_{i}\, \lambda_i \lmlm{i}{i}$.
For type III and $\mathrm{III}_i$ tensors, the generic situation is
$C_{011i} \neq 0$, and thus the tensor defines a preferred null plane
spanned by $C_{011i} \vm^i$ and $\vell$.  One can therefore put the tensor
into canonical form by performing a spin and a boost so that
$C_{0112}=1$ and $C_{011i}=0$ for all $i=3,4,\ldots$.


{\bf 4. Riemann-like tensors.}  Much of the analysis for higher
dimensional tensors can be applied directly to the classification of
higher dimensional Riemann curvature tensors.  In particular the
higher-dimensional alignment types shown in Table \ref{tab:weylclass}
give well defined categories for the Riemann tensor.  However, in the
case of the Riemann tensor there are additional constraints coming
from the extra non-vanishing components.  For example, a type I
Riemann curvature tensor must satisfy the $\frac12 m(m+1)$ constraints
$R_{0i0j}=0$, whereas a type I Weyl-like tensor has one less constraint
owing to the fact that $C_{0i0}{}^i=0$.


{\bf 5. Differential-geometric considerations.}  The focus of the
present paper has been purely algebraic.  We have been studying
tensors rather than tensor fields, or to put it another way we have
been considering tensors at single point of a Lorentzian manifold.
There is however, a rich interplay between algebraic type and the
differential Bianchi identities.

In 4 dimensions, there are a number of consequences, such as the
Goldberg-Sachs theorem and its non-vacuum generalizations
\cite{kramer,penrind,goldsachs}.  In higher dimensions, differential
consequences of the Bianchi identities in type III and N spacetimes
have been considered in \cite{ppcm}.

In 4D it is possible to use the Bianchi and Ricci equations to
construct many algebraically special solutions of Einstein's field
equations. The hope is that it is possible to do a similar thing in
higher dimensions, at least for the simplest algebraically special
spacetimes. The vast majority of today's known higher-dimensional
exact solutions are simple generalizations of 4D solutions. This
approach may lead to new, genuinely higher dimensional exact
solutions. Type N and D solutions may be of particular physical
interest.  The Goldberg-Sachs type theorems are very useful for
constructing algebraically special exact solutions. At present it is
unclear to what extent such theorems may be generalized for higher
dimensions.

One further application is the classification of VSI manifolds in
higher dimensions.  The primary alignment type of such manifolds would
have to be III, N, or  O, while the local coordinate expression for
the metric would be a higher-dimensional generalization of the Kundt
form\cite{kramer,cmpppz,gvsi}.

\begin{acknowledgements}
  RM and AC were partially supported by a research grant from NSERC.
  VP was supported by GACR-202/03/P017 We would like to thank Nicos
  Pelavas for helpful comments.  AP and VP would like to thank
  Dalhousie University for its hospitality while this work was carried
  out.  
\end{acknowledgements}

\bibliographystyle{apsrev}
\bibliography{algspec}
\end{document}